\documentclass[12pt,preprint]{aastex} 

\usepackage{amsmath,amssymb}
\usepackage{graphicx}
\usepackage{float}

\slugcomment{KSUPT-12/3 \hspace{0.5truecm} June 2012}

\begin{document}
\title{Forecasting cosmological parameter constraints from near-future space-based galaxy surveys}

\author{Anatoly Pavlov\altaffilmark{1}, Lado Samushia\altaffilmark{2}, and Bharat Ratra\altaffilmark{1}}

\altaffiltext{1}{Department of Physics, Kansas State University, 
                 116 Cardwell Hall, Manhattan, KS 66506, USA \ 
                 pavlov@phys.ksu.edu, ratra@phys.ksu.edu}
\altaffiltext{2}{Institute of Cosmology and Gravitation, University of
                 Portsmouth, Dennis Sciama Building, Portsmouth, P01 3FX, UK \
                 lado.samushia@port.ac.uk}

\begin{abstract}
The next generation of space-based galaxy surveys are expected to
measure the growth rate of structure to about a percent level over a range of
redshifts. The rate of growth of structure as a function of redshift depends on the behavior of dark energy and so can be used to constrain parameters of 
dark energy models. In this work we investigate how well these future data 
will be able to constrain the time dependence of the dark energy density. 
We consider parameterizations of the dark energy equation of state, such as 
XCDM and $\omega$CDM, as well as a consistent physical model of time-evolving 
scalar field dark energy, $\phi$CDM. We show that if the standard, specially-flat cosmological model is taken as a fiducial model of the Universe, these near-future measurements of structure growth will be able to constrain the time-dependence of scalar 
field dark energy density to a precision of about 10\%, which is almost an 
order of magnitude better than what can be achieved from a compilation of 
currently available data sets.
\end{abstract}

\section{Introduction}

Recent measurements of the apparent magnitude of Type Ia supernovae (SNeIa)
continue to indicate, quite convincingly, that the cosmological expansion 
is currently accelerating \citep[see, e.g.,][]{Conley:2011ab, Suzuki:2012ab, 
Li:2011ab, Barreira:2011ab}.

If we assume that general relativity provides an adequate description of
gravitational interactions on these cosmological length scales, then the 
kinematic properties of the Universe can be derived by solving the 
Einstein equations
\begin{equation}
   R_{\mu \nu} - \frac{1}{2}g_{\mu \nu}R = 8G\pi T_{\mu \nu}.
 \label{equ:en}
\end{equation}
\noindent
Here $g_{\mu\nu}$ is the metric tensor, $R_{\mu\nu}$ and $R$ are the Ricci 
tensor and (curvature) scalar respectively, $T_{\mu\nu}$ is the stress-energy tensor of
the Universe's constituents, and $G$ is the Newtonian gravitational constant.

There is good observational evidence that the large-scale radiation and 
matter distributions are statistically spatially isotropic. The (Copernican)
cosmological principle, which is also consistent with current observations,
then indicates that the Friedmann-Lema{\^\i}tre-Robertson-Walker (FLRW) 
models provide an adequate description of the spatially homogeneous background
cosmological model. 

In the FLRW models, the current accelerating cosmological 
expansion is a consequence of dark energy, the dominant, by far, term 
in the current cosmological energy budget. The dark energy density could 
be constant in time (and hence uniform in space) --- Einstein's cosmological 
constant $\Lambda$ \citep{Peebles:1984ab} --- or gradually decreasing in 
time and thus slowly varying in space \citep{Peebles&Ratra:1988ab}.
 
The ``standard" model of cosmology is the spatially-flat $\Lambda$CDM 
model in which the cosmological constant contributes around 75\% of the
current energy budget. Non-relativistic cold dark matter (CDM) is the 
next largest contributor, at around 20\%, with non-relativistic baryons
in third place with about 5\%. For a review of the standard model see
\citet{Ratra:2008ab} and references therein. 

Recent measurements of the anisotropies of the cosmic microwave background
(CMB) radiation \citep[e.g.,][]{Komatsu:2011ab, Reichardt:2012ab}, in 
conjunction with significant observational support for a low density 
of non-relativistic matter 
\citep[CDM and baryons together, e.g.,][]{Chen:2003ab}, as well as 
measurements of the position of the baryon acoustic oscillation (BAO) 
peak in the matter power spectrum \citep[e.g.,][]{Percival:2010ab, 
Dantas:2011ab, Carnero:2012ab, Anderson:2012ab}, provide significant 
observational support to the 
spatially-flat $\Lambda$CDM model. Other data are also not inconsistent
with the standard $\Lambda$CDM model. These include strong gravitational 
lensing measurements \citep[e.g.,][]{Chae:2004ab, Lee:2007ab, 
Biesiada:2010ab}, measurement of Hubble parameter as a function of redshift 
\citep[e.g.,][]{Samushia:2006ab, Sen:2008ab, Pan:2010ab, Chen:2011ab}, 
large-scale structure data \citep[e.g.,][]{Baldi:2011ab, DeBoni:2011ab, 
Brouzakis:2011ab, Campanelli:2011ab}, and galaxy cluster gas mass fraction 
measurements \citep[e.g.,][]{Allen:2008ab, Samushia:2008ab, Tong:2011ab}.
For recent reviews of the situation 
see, e.g., \citet[]{Blanchard:2010ab}, \citet[]{Sapone:2010ab}, and 
\citet[]{Jimenez:2011ab}.

While the predictions of the $\Lambda$CDM model are in reasonable accord
with current observations, it is important to bear in mind that 
dark energy has not been directly detected (and neither has dark matter).
Perhaps as a result of this, some feel that it is more reasonable to assume
that the left hand side of Einstein's Eq.~(\ref{equ:en}) needs to be 
modified (instead of postulating a new, dark energy, component of the 
stress-energy tensor on the right hand side). While such modified gravity 
models are under active investigation, at present there is no compelling 
observational reason to prefer any of these over the standard 
$\Lambda$CDM cosmological model.

The $\Lambda$CDM model assumes that dark energy is a cosmological constant
with equation of state
\begin{equation}
  p_\Lambda = -\rho_\Lambda,
  \label{equ:eos}
\end{equation}
\noindent
where $p_\Lambda$ and $\rho_\Lambda$ are the pressure and energy density of 
the cosmological constant (fluid). This minimalistic model, despite being in 
good agreement with most observations available today, has some potential 
conceptual shortcomings that have prompted research into alternative 
explanations of the dark energy phenomenon.\footnote{
Structure formation in the $\Lambda$CDM model is governed by the 
``standard" CDM structure formation model, which might be in 
some observational difficulty \citep[see, e.g.,][]{Peebles:2003ab,
Perivolaropoulos:2010ab}.} 

To describe possible time-dependence of the dark energy density, it 
has become popular to consider a more general equation of state 
parametrization
\begin{equation}
  p_\omega = \omega(z)\rho_\omega.
  \label{equ:wrho}
\end{equation}
\noindent
Here $p_\omega$ and $\rho_\omega$ are the pressure and energy density of 
the dark energy fluid with redshift $z$ dependent equation of state 
parameter $\omega(z)$. The simplest such parametrization is the XCDM one
for which the equation of state parameter is constant and results in accelerated expansion if 
$\omega(z) = \omega_X < -1/3$. In this case the dark energy density 
decreases with time and this allows for the possibility that the 
fundamental energy density scale for dark energy is set at high 
energy in the early Universe and the slow decrease of the energy 
density over the long age of the Universe ensures that the characteristic 
dark energy density scale now is small (a few meV). This also ensures that
the dark energy density remains comparable to the matter energy 
density over a longer period of time (compared to that for the 
$\Lambda$CDM model).

When $\omega_X = -1$ the XCDM parametrization reduces to the consistent
(and complete) $\Lambda$CDM model. For any other value of $\omega_X$ the XCDM parametrization cannot consistently describe
spatial inhomogeneities without further assumptions and extension
\citep[see, e.g.,][]{Ratra:1991ab, Podariu:2000ab}. Models in which
 $\omega(z)$ varies in time, $\omega$CDM models, are also unable
to consistently describe spatial inhomogeneities without further 
assumptions and extension.

A physically and observationally viable alternative to the 
$\Lambda$CDM model, that consistently describes a slowly decreasing
in time dark energy density, is the $\phi$CDM model 
\citep{Peebles&Ratra:1988ab, Ratra:1988ab}. This model, in which a 
dark energy scalar field, $\phi$, slowly roles down its potential, 
resulting in a slowly decreasing dark energy 
density, alleviates some of the conceptual problems, mentioned above, 
associated with the $\Lambda$CDM model. The slowly rolling scalar 
field, at a given instant of time, can be approximated by a dark
energy fluid with an appropriately negative equation of state parameter.

More specifically, a $\phi$CDM model with an inverse-power-law scalar field
potential energy density $V(\phi)\propto \phi^{-\alpha}$, $\alpha > 0$, is 
a prototypical example that has been extensively studied. This model has a 
non-linear attractor or ``tracker" scalar field solution that forces the 
initially sub-dominant dark energy density to come to dominate over the matter
energy density, thus dominating the energy budget of the current Universe, 
and so resulting in the current accelerated cosmological expansion. In 
addition to therefore partially alleviating the ``coincidence" problem of the 
$\Lambda$CDM model, the $\phi$CDM model generates the current tiny dark 
energy scale of order an meV, measured by the SNeIa, through decrease,
via cosmological expansion over the long age of the Universe, of a much larger
energy scale.  

The $\alpha$ parameter controls the steepness of the scalar field 
potential, with larger values resulting in a stronger time dependence 
of the approximate equation of state parameter and $\alpha=0$ 
corresponds to the $\Lambda$CDM model limit. $\alpha$ has been
constrained using currently available data \citep[see e.g.,][and references
therein]{Chen:2004ab, Wilson:2006ab, Chen:2012ab, Mania:2012ab}. 
The strongest current limits are that $\alpha$ has to be less than $\sim 0.7$ 
at 2$\sigma$ confidence \citep{SamushiaThesis}.

In the $\phi$CDM model, or in the XCDM or $\omega$CDM parameterizations, 
the background evolution of the (spatially homogeneous) Universe differs
from that in the $\Lambda$CDM case. This affects both the distance-redshift
relation as well as the growth rate of large-scale structure. With precise 
measurements of distance and growth rate over a range of redshifts it will 
be possible to discriminate between cosmological models.\footnote{
There are many other models under current discussion, besides the 
$\Lambda$CDM and $\phi$CDM models and XCDM and $\omega$CDM parameterizations 
we consider here for illustrative purposes. For a sample of the available
options see, e.g., \citet{Yang:2011ab}, \citet{Frolov:2011ab}, 
\citet{Nunes:2011ab}, \citet{Grande:2011ab}, \citet{Saitou:2011ab}, 
\citet{Silva:2012ab}, \citet{Kamenshchik:2011ab}, and \citet{Maggiore:2011ab}.}

The BAO signature in the observed large-scale structure of the Universe allows
for the measurements of radial and angular distances as functions of redshift 
\citep[see, e.g.,][]{Percival:2010ab, Blake2011b, Beutler2011, 
Anderson:2012ab}.
In addition, the redshift-space distortion signal allows for inferences about 
the strength of gravitational interactions on very large scales \citep[see,
e.g.,][]{Percival2004, Angulo2008, Guzzo2008, Blake2011a, Samushia2012, 
Reid:2012ab}. Currently available data sets have been used to measure 
distances and growth history up to a redshift $z \sim 0.8$ and the next 
generation of planned space-based galaxy redshift surveys of the whole 
extragalactic sky are expected to extend these measurements to a redshift 
$z \sim 2$.  Possible candidates for such surveys include the Euclid 
satellite mission that has been approved by the European Space Agency 
\citep{Laureijs2011} and the WFIRST satellite that was ranked high by the 
recent Decadal Survey \citep{Green2011}. These surveys have been shown to 
have the potential to measure angular distances, Hubble parameter $H(z)$, 
and growth rate as functions of redshift to a few percent precision over 
a wide range of redshifts \citep[]{Wang2011, samushia2010, 
Majerotto2012, Basse:2012}.\footnote{
For constraints on cosmological parameters from data from space missions
proposed earlier, see \citet{Podariu:2001ab} and references therein.} 

As mentioned above, an alternative potential explanation of the observed 
accelerated expansion of the Universe is to replace general relativity
by a modified theory of gravity. 
For example, in the $f(R)$-gravity models the Einstein-Hilbert gravitational 
action is modified to
\begin{equation}
S = \frac{1}{16\pi G}\int d^{4}x\sqrt{-g}f(R),
\label{equ:Act_fR}
\end{equation}
where the function $f(R)$ of the Ricci curvature $R$ can in general be 
of any form. In the special case when $f(R) = R$ one recovers the 
Einstein-Hilbert action which yields the Einstein equations of 
general relativity, Eq.\ (\ref{equ:en}). For every dark energy model it is 
possible to find a function $f(R)$
that will result in exactly the same expansion history
\citep[see, e.g.][]{Sotiriou:2010ab, Tsujikawa:2010ab, Capazziello2011ab} 
 thus potentially eliminating the need for dark energy. However, nothing 
prevents the coexistence of modified gravity 
and dark energy, with both contributing to powering the current accelerated 
cosmological expansion. It is of significant importance to be able to 
determine which scenario best describes what is taking place in our 
Universe.

In this paper we investigate how well anticipated data from the galaxy
surveys mentioned above can constrain the time dependence of the dark energy. 
We will use the Fisher matrix formalism to obtain predictions for the 
$\phi$CDM model and compare these with those made using the (model-dependent) 
XCDM and $\omega$CDM parameterizations of dark energy. We will mostly
assume that gravity is well described by general relativity, but will also
look at some simple modified gravity cases. We find that the anticipated 
constraints on the parameter $\alpha$ of the $\phi$CDM model are almost an order of 
magnitude better than the ones that are currently available.

Compared to the recent analysis of \citet{samushia2010}, here we use 
an updated characterization of planned next-generation space-based galaxy 
surveys, so our forecasts are a little more realistic. We also consider an
additional dark energy parametrization, XCDM, a special case of $\omega$CDM 
that was considered by \citet{samushia2010}, as well as the $\phi$CDM
model, forecasting for which has not previously been done.

The paper is organized as follows. In Sec.\ \ref{sec:measurement} we briefly 
describe the observables and their relationship to basic cosmological 
parameters. In Sec.\ \ref{sec:model} we describe the models of dark energy 
that we study. Section \ref{sec:fisher} outlines the method we use for 
predicting parameter constraints, with some details given in the Appendix. 
We present our results in Sec.\ \ref{sec:result} and conclude in 
Sec.\ \ref{sec:conclusion}.

\section{Measured power spectrum of galaxies} 
\label{sec:measurement}

The large-scale structure of the Universe, which most likely originated as 
quantum-mechanical fluctuations of the scalar field that drove an early 
epoch of inflation \citep[see, e.g.,][]{Fischler:1985ab}, became 
(electromagnetically) observable at $z \sim 10^{3}$ after the recombination 
epoch. Dark
energy did not play a significant role at this early recombination epoch 
because of its low mass-density relative to the densities of ordinary and 
dark matter as well as that of radiation (neutrinos and photons). At 
$z \sim$ 5 galaxy clusters began to form. Initially, in regions where the 
matter density was a bit higher than the average, space expanded a bit 
slower than average. Eventually the dark and ordinary matter reached a 
minimum density and the regions contracted. If an over-dense region was 
sufficiently large its baryonic matter collapsed into its dark-matter halo. 
The baryonic matter continued to contract even more due to its ability 
to lose thermal energy through the emission of electromagnetic radiation.
This can not happen with dark matter since it does not emit significant 
electromagnetic radiation nor does it interact significantly 
(non-gravitationally) with 
baryonic matter. As a result the dark matter remained in the form of 
a spherical halo around the rest of the baryonic part of a galaxy. At 
$z \sim$ 2 the rich clusters of galaxies were formed by gravity, which 
gathered near-by galaxies together. Also by this time the dark energy's 
energy density had become relatively large enough to affect the growth of 
large-scale structure.

Different cosmological models with different sets of parameters 
can result in the same expansion history and so it impossible to distinguish 
between such models by using only expansion history measurements. This is 
one place where measurements of the growth history of the large-scale 
structure of the Universe plays an important role. It is not possible to fix free parameters
of two different cosmological models to give exactly the same expansion 
and growth histories simultaneously. It is therefore vital to observe both histories in 
order to obtain better constraints on parameters of a cosmological model.

In a cosmological model described by the FLRW metric, and to lowest 
order in dark matter over-density perturbations, the power-spectrum of
observed galaxies is given by \citep{kaiser87}
\begin{equation} P_{g}(k,\mu) = P_{m}(k)(b\sigma_{8} + f\sigma_{8}\mu^{2})^{2}.
  \label{equ:psigma} 
\end{equation} 
\noindent  
Here subscript $g$ denotes galaxies, $P_m$ is the underlying matter power
spectrum, $b$ is the bias of galaxies, $f$ is the growth rate, $\mu$ is 
the cosine of the angle between wave-vector $k$ and the line-of-sight 
direction, and $\sigma_8$ is the overall normalization of the power spectrum
($\sigma_8$ is the rms energy density perturbation smoothed over spheres of 
radius $8 h^{-1}$ Mpc, where $h = H_0/(100 {\rm km} {\rm s}^{-1} 
{\rm Mpc}^{-1}$) and $H_0$ is the Hubble constant). 
Since, for a measured power spectrum of galaxies on a single redshift slice, 
the bias and growth rate are perfectly degenerate with the overall amplitude, 
in the equations below we will refer to $b\sigma_8$ and $f\sigma_8$ simply as
$b$ and $f$.  

The angular dependence of the power spectrum in Eq.\ (\ref{equ:psigma}) can be 
used to infer the growth rate factor $f(z)$ which is defined as the logarithmic derivative of 
the linear growth factor
\begin{equation}
f(z) = \frac{d \ln G}{d \ln a},
 \label{equ:fz_dlg}
\end{equation}
\noindent
where $a$ is the cosmological scale factor, and the linear growth factor 
$G(t) = \delta(t)/\delta(t_{\rm in})$ shows by how much the perturbations 
have grown since some initial time $t_{\rm in}$.\footnote{
Here we have expanded the energy density $\rho (t,{\vec x})$ in terms of a 
small spatially inhomogeneous fractional perturbation $\delta (t, {\vec x})$
about a spatially-homogeneous background $\rho_b(t)$: $\rho (t,{\vec x}) =
\rho_b(t) [ 1 + \delta (t, {\vec x})]$.}

The numerical value of the $f(z)$ function depends both on the theory of
gravity and on the expansion rate of the Universe. Since the growth rate 
depends very sensitively on the total amount of non-relativistic matter, 
it is often parametrized as 
\citep[see, e.g.,][and references therein]{linder05} 
\begin{equation}
f(z) \approx \Omega_{m}^{\gamma}(z), 
 \label{equ:fzOm}
\end{equation}
\noindent
where
\begin{equation}
\Omega_{m}(z) = \frac{\Omega_{m}(1 + z)^{3}}{E^2(z)},
\label{equ:Omz}
\end{equation}
\noindent
and 
\begin{equation}
E(z) = H(z)/H_0 = \sqrt{\Omega_{m}(1 + z)^{3} + \Omega_{k}(1 + z)^{2} + 
       \Omega_{DE}(z)}.
 \label{equ:Ez}
\end{equation}
\\
Here $H(z)$ is the Hubble parameter and $H_0$ is its value at the present
epoch (the Hubble constant), $\Omega_{m}$ is the value of the energy density 
parameter of non-relativistic matter at the present epoch ($z = 0$), 
$\Omega_k$ that of spatial curvature, and $\Omega_{DE}(z)$ is the energy
density parameter which
describes the evolution of the dark energy density and is different in 
different dark energy models. 

The growth index, $\gamma$, depends on both a model of dark energy as well 
as a theory of gravity. When general relativity is assumed and the equation
of state of dark energy is taken to be of the general form in Eq.\ 
(\ref{equ:wrho}) then 
\citep[see, e.g.,][and references therein]{linder05}
\begin{equation}
\gamma \approx 0.55 + 0.05[1 + \omega(z=1)]
 \label{equ:gamma}
\end{equation}
\noindent
to a few percent accuracy.
In the $\Lambda$CDM cosmological model $\gamma \approx$ 0.55. An observed 
significant deviation from this value of $\gamma$ will present a 
serious challenge for the standard cosmological model.

The power spectrum is measured under the assumption of a fiducial 
cosmological model. If the angular and radial distances in the 
fiducial model differ from those in the real cosmology, the power spectrum 
will acquire an additional angular dependence via the \citet[][AP]{alcock79} 
effect, as discussed in \citet{samushia2010}, 
\begin{equation}
P_{g}(k, \mu) = \frac{1}{f_{\parallel}f_{\perp}^{2}}
                 P_{m}\left(\frac{k}{f_{\perp} F}\sqrt{F^2 + 
                 \mu^{2}\left(1 - F^2\right)} \right)
                 \times\left\{b + \frac{\mu^{2}f}{F^{2} + 
                 \mu^{2}(1 - F^{2})}\right\}^{2}, 
\label{equ:Pw}
\end{equation}
\noindent
where
\begin{equation}
f_{\parallel}(z) = R_{r}(z)/\hat{R}_{r}(z),
 \label{equ:f_par}
\end{equation}
\begin{equation}
f_{\perp}(z) = D_{A}(z)/\hat{D}_{A}(z),
 \label{equ:f_per}
\end{equation}
\begin{equation}
F = f_{\parallel}/f_{\perp}.
\end{equation}
\noindent
Here $R_{r} = dr/dz$ is the derivative of the radial distance, $D_{A}$ is 
the angular diameter distance (both defined below), a hat indicates a quantity 
evaluated in the fiducial cosmological model, and a quantity without a hat is 
evaluated using the alternative cosmological model. The AP effect is an 
additional source of anisotropy in the measured power spectrum and allows 
for the derivation of stronger constraints on cosmological parameters.

\section{Cosmological models}
\label{sec:model}

In an FLRW model with only non-relativistic matter and dark energy the 
distances $D_{A}(z)$ and $R_{r}(z)$ are 
\begin{equation}
D_{A}(z) = \frac{1}{h\sqrt{\Omega_{k}}(1 + z)}
           \sinh\left( \sqrt{\Omega_{k}}\int_{0}^{z}\frac{dz'}{E(z')} \right),
\label{equ:Dz}
\end{equation}
\noindent
\begin{equation}
R_{r}(z) = \frac{1}{h(1 + z)E(z)}.
\label{equ:Rr}
\end{equation}
Here $E(z)$ is defined in Eq.\ (\ref{equ:Ez}). The
functional form of $E(z)$ depends on the model of dark energy.

\subsection{$\Lambda$CDM, XCDM and $\omega$CDM parameterizations}

Here we describe the relevant features of the $\Lambda$CDM model and the 
dark energy parameterizations we consider.

If the dark energy is taken to be a fluid its equation of state can be 
written as $p = \omega(z)\rho$. For the $\Lambda$CDM model the equation
of state parameter $\omega(z) = -1$ and the dark energy density is time 
independent.

In the XCDM parametrization $\omega (z) = \omega_X (< -1/3)$ is allowed 
to take any time-independent value, resulting in a time-dependent dark 
energy density.

In the $\omega$CDM parametrization the time dependence of $\omega(z)$ is
parametrized by introducing an additional parameter $\omega_a$ through
\citep[][]{Chevallier:2001ab, Linder:2003ab}
\begin{equation}
w(z) = w_{0} + w_{a}\frac{z}{1 + z}.
 \label{equ:wz}
\end{equation}
\noindent
The XCDM parametrization is the limit of the $\omega$CDM parametrization 
with $\omega_a = 0$. In the $\omega$CDM parametrization the function 
$\Omega_{DE}(z)$ that describes the time evolution of the dark energy
density is 
\begin{equation}
\Omega_{DE}(z) = (1 - \Omega_{m} - \Omega_{k}) 
      (1 + z)^{3(1 + w_{0} + w_{a})}\exp\left(-3w_{a}\frac{z}{1 + z}\right), 
 \label{equ:Fz}
\end{equation}
\noindent
and the corresponding expression for the XCDM case can be derived by setting 
$\omega_a = 0$ here.

\subsection{$\phi$CDM model}

In the $\phi$CDM model the energy density of the background, spatially 
homogeneous, scalar field $\phi$ can be found by solving 
the set of simultaneous ordinary differential equations of motion, 
\begin{equation}
\ddot{\phi} + 3\frac{\dot{a}}{a}\dot{\phi} + V'(\phi) = 0
 \label{equ:phi_field},
\end{equation}
\begin{equation}
\left(\frac{\dot{a}}{a} \right)^{2} = \frac{8\pi G}{3}(\rho + \rho_{\phi}) - \frac{k}{a^{2}},
 \label{equ:phi_hubble}
\end{equation}
\begin{equation}
\rho_{\phi} = \frac{1}{16\pi G}\left(\frac{1}{2}\dot{\phi}^{2} 
   + V(\phi) \right).
 \label{equ:phi_rho}
\end{equation}
\noindent
Here an over-dot denotes a derivative with respect to time, a prime denotes
one with respect to $\phi$, $V(\phi)$ is the potential energy density of the
scalar field, $\rho_\phi$ is the energy density of the scalar field, and 
$\rho$ that of the other constituents of the Universe.

Following \citet{Peebles&Ratra:1988ab} we consider a scalar field with 
inverse-power-law potential energy density
\begin{equation}
V(\phi) = \frac{\kappa}{2G}\phi ^{-\alpha}.
 \label{equ:phi_V}
\end{equation}
\noindent
Here $\alpha$ is a positive parameter of the model to be determined
experimentally and $\kappa$ is a positive constant. This choice of potential
has the interesting property that the scalar field solution is an attractor 
with an energy density that slowly comes to dominate over the energy 
density of the non-relativistic matter (in the matter dominated epoch) and 
causes the cosmological expansion to accelerate. The function 
$\Omega_{DE}(z)$ in the case of $\phi$CDM is
\begin{equation}
\Omega_{DE}(z) = \frac{1}{12}\left(\dot{\phi}^{2} 
   + \frac{\kappa}{G}\phi ^{-\alpha} \right).
 \label{equ:phi_Ode}
\end{equation}

\section{Fisher matrix formalism}
\label{sec:fisher}

The precision of the galaxy power spectrum measured in redshift bins 
depends on the cosmological model, the volume of the survey, and the 
distribution of galaxies within the observed volume. See App.\ A for 
a summary of how to estimate the precision of measurements from 
survey parameters.

We assume that the power spectrum $P(k_i)^{\rm meas}$ has been measured in 
$N$ wave-number $k_i$ bins ($i = 1\ldots N$) and each measurement has a 
Gaussian uncertainty $\sigma_{i}$.  From these measurements a likelihood function
\begin{equation}
\mathcal{L} \propto \exp\left(-\frac{1}{2}\chi^{2} \right)
 \label{equ:likelihood}
\end{equation}
\noindent
can be constructed where
\begin{equation}
  \chi^{2} = \sum_{i = 1}^{N}\frac{(P_{i}^{\rm meas} - 
    P_{i}(\vec{p}))^{2}}{\sigma_{i}^{2}} .
 \label{equ:chi}
\end{equation}
\noindent
Here $\vec{p}$ are the set of cosmological parameters on which 
the power spectrum depends.

The likelihood function in Eq.~(\ref{equ:likelihood}) can be transformed into
the likelihood of theoretical parameters $\vec{p}$ by Taylor expanding it
around the maximum and keeping terms of only second order in 
$\delta \vec{p}$ as $\chi^{2}(\delta p)$ = $ F_{jk} \delta p^{j} 
\delta p^{k}$, where $F_{jk}$ is the
Fisher matrix\footnote{
For a review of the Fisher matrix formalism as applied to cosmological
forecasting, see \citet[]{albrecht09}.}
of the parameter set $\vec{p}$ given by second derivatives of the
likelihood function through
\begin{equation}
F_{jk} = -\left\langle \frac{\partial^{2}\ln \mathcal{L}}{\partial p^{j}\partial p^{k}} 
         \right\rangle .
 \label{equ:F_jk}
\end{equation}

The Fisher matrix predictions are exact in the limit where initial measurements
as well as derived parameters are realizations of a Gaussian random variable.
This would be the case if the $P_{i}^{\rm meas}$ were perfectly Gaussian and 
the $P_{i}(\vec{p})$ were linear functions of $\vec{p}$, which would make the 
second order Taylor expansion of the likelihood around its best fit value 
exact. In reality, because of initial non-Gaussian contributions and 
nonlinear effects, the predictions of Fisher matrix analysis will be 
different (more optimistic) from what is achievable in practice. These 
differences are larger for strongly non-linear models and for the phase 
spaces in which the likelihood is non-negligible at some physical boundary 
($\alpha=0$ in case of $\phi$CDM). A more realistic approach, that requires 
significantly more computational time and power, is to generate a large 
amount of mock data and perform a full Monte-Carlo Markov Chain
(MCMC) analysis \citep[see, e.g.,][where the authors find significant
differences compared to the results of the Fisher matrix
analysis]{Perotto2006,Martinelli2011}.

We assume that the full-sky space-based survey will observe H$\alpha$-emitter
galaxies over 15000 $\rm{deg}^2$ of the sky. For the density and bias of
observed galaxies we use predictions from \citet{Orsi:2010ab} and
\citet{Geach:2010ab} respectively. We further assume that about half of the
galaxies will be detected with a reliable redshift. These numbers roughly mirror
what proposed space missions, such as the ESA Euclid satellite and the NASA
WFIRST mission, are anticipated to achieve.  For the fiducial cosmology we use a
spatially-flat $\Lambda$CDM model with $\Omega_{m}$ = 0.25, the baryonic matter
density parameter $\Omega_{b}$ = 0.05, $\sigma_{8}$ = 0.8, and the primordial
density perturbation power spectral index $n_{s}$ = 1.0, for convenience we
summarize all the parameters of the fiducial model in Table 1.

\begin{table}[H]
\caption{Values of the parameters of the fiducial $\Lambda$CDM model and the survey.}
\centering
\begin{tabular}{|c|c|c|c|c|c|c|c|c|}
  \hline
  $\Omega_{m}$ & $\Omega_{b}$ & $\Omega_{k}$ & h & $\sigma_{8}$ & $n_{s}$  & Efficiency  & Redshift span & Covered sky area in $\rm{deg}^2$ \\
  \hline
  0.25 & 0.05 & 0.0 & 0.7 & 0.8 & 1.0 & 0.45 & 0.55 $\leq z \leq$ 2.05 & 15000 \\
  \hline
\end{tabular}
 \label{tab: fiducial}
\end{table}

We further assume that the shape of the power spectrum is known perfectly (for
example from the results of the {\it Planck} satellite) and ignore derivatives
of the real-space power spectrum with respect to cosmological parameters.

We predict the precision of the measured galaxy power spectrum and then 
transform it into correlated error bars on the derived cosmological 
parameters. At first we make predictions for the basic quantities $b$ and 
$f$ in the XCDM and $\omega$CDM parameterizations and in the $\phi$CDM model.
Then it allows us to predict constraints on deviations from general relativity and see 
how these results change with changing assumptions about dark energy. 
Finally, we forecast constraints on the basic cosmological parameters of 
dark energy models.  

For the XCDM parametrization these basic cosmological
parameters are $p_{\rm XCDM}=(f, b, h, \Omega_m, \Omega_k, w_X)$. The 
$\omega$CDM parametrization has one extra parameter describing the time
evolution of the dark energy equation of state parameter, $p_{\omega {\rm CDM}}
= (f, b, h, \Omega_m, \Omega_k, w_0, w_a)$. For the $\phi$CDM model the
time dependence of the dark energy density depends only on one parameter 
$\alpha$ so we have $p_{\phi{\rm CDM}}=(f, b, h, \Omega_m, \Omega_k, \alpha)$. 
In order to derive constraints on the parameters of the considered 
cosmological models while altering assumptions about the correctness of 
general relativity, we transform Fisher matrices of each model from the 
parameter set described above to the following parameter set (that 
now includes $\gamma$ that parametrizes the growth rate) $p_{model} = 
(\gamma, model)$, where by $model$ we mean all the parameters of a 
particular model, for example, for $\omega$CDM $model 
= p_{\omega {\rm CDM}} = (f, b, h, \Omega_m, \Omega_k, w_0, w_a)$.

\section{Results}
\label{sec:result}

\subsection{Constraints on growth rate}

\begin{figure}[H]
   \centering
   \includegraphics[width=1.0\textwidth]{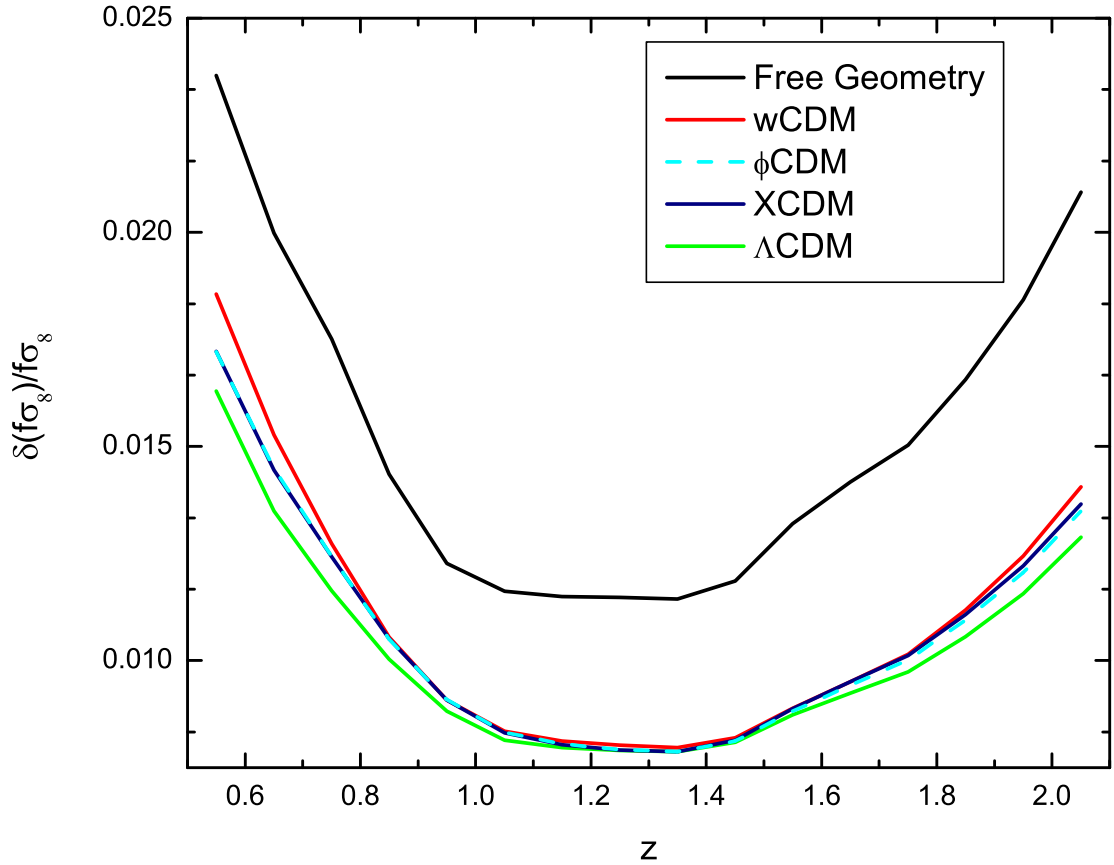}
    \caption{Predicted relative error on the measurements of growth rate as a 
     function of redshift $z$ in redshift bins of $\Delta z = 0.1$ for 
     different models of dark energy. The upper solid black line shows 
     predictions for the case when no assumption is made about the nature of 
     dark energy.}
 \label{fig: 1}
\end{figure}

Figure \ref{fig: 1} shows predictions for the measurement of growth rate 
assuming different dark energy models. We find that in the most general 
case, when no assumption is made about the nature of dark energy, the 
growth rate can be constrained to a precision of better then 2\% over a
wide range of redshifts. This is in good agreement with previous similar 
studies \citep[see, e.g., Fig.\ 1 of][]{samushia2010}. When we specify 
a dark energy model the constraints on growth rate improve by about a 
factor of two. There is very little difference between the results derived 
for different dark energy models: the precision is almost insensitive to 
the assumed model. Also, one can notice that the curves for the XCDM 
parametrization and for the $\phi$CDM model are almost identical. The 
likely explanation of this effect is that for a fixed redshift bin the 
$\phi$CDM model is well described by the XCDM parametrization with the 
value of the parameter $\omega_{X}$ = $p_{\phi}/\rho_{\phi}$, where the 
values of the scalar field pressure $p_{\phi}$ and energy density 
$\rho_{\phi}$ are evaluated at that redshift bin. 

\begin{figure}[H]
   \centering
   \includegraphics[width=1.0\textwidth]{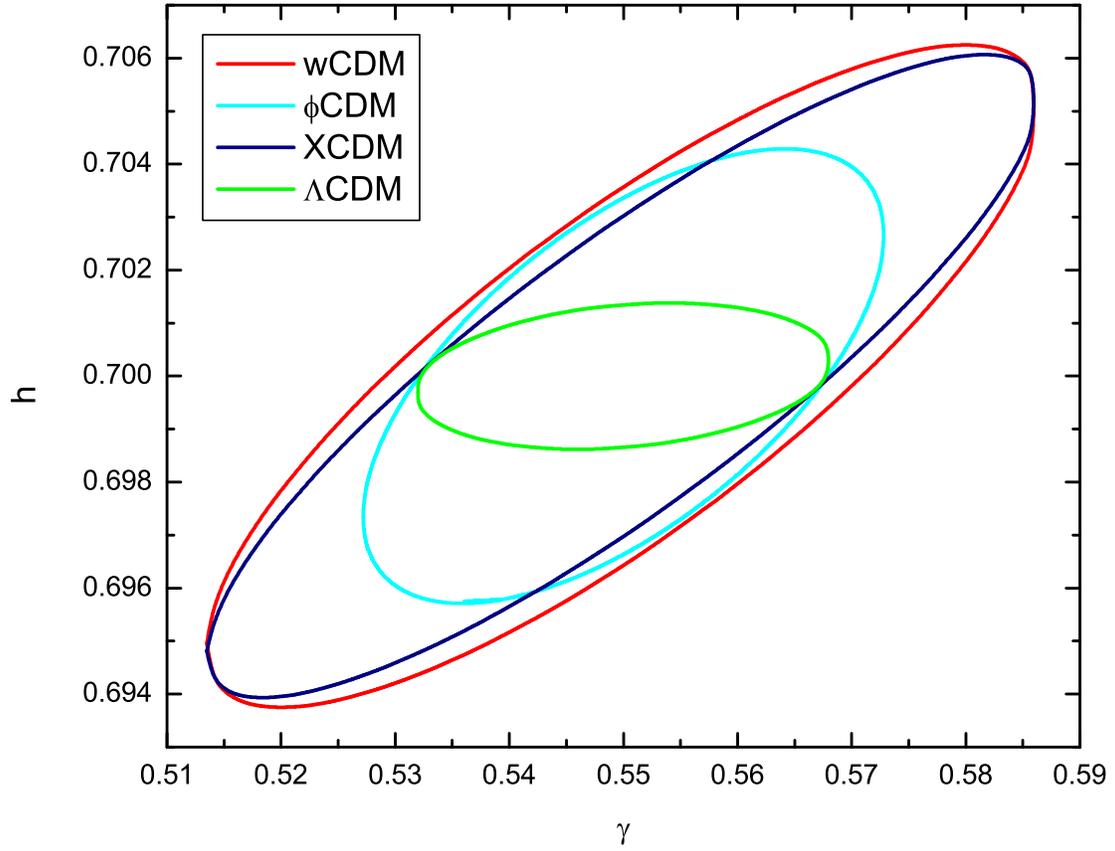}
    \caption{Predicted one standard deviation confidence level contour 
     constraints on 
     the current renormalized Hubble constant $h$ and the parameter $\gamma$ 
     that describes deviations from general relativity for different dark
     energy models.}
 \label{fig: 2}
\end{figure}

The measurements of growth rate can be remapped into constraints on 
parameters describing the deviation from general relativity. Figure 
\ref{fig: 2} shows correlated constraints between the current 
re-normalized Hubble constant $h$ and the $\gamma$ parameter that describes
the growth of structure. The $\phi$CDM model constraints on both $h$
and $\gamma$ are tighter than those for the XCDM or $\omega$CDM 
parameterizations. As expected, the most restrictive $\Lambda$CDM model 
results in the tightest constraints.

\subsection{Constraints on dark energy model parameters}

We use measurements of growth and distance to constrain parameters of the 
dark energy models. 

\begin{figure}[H]
   \centering
   \includegraphics[width=0.75\textwidth]{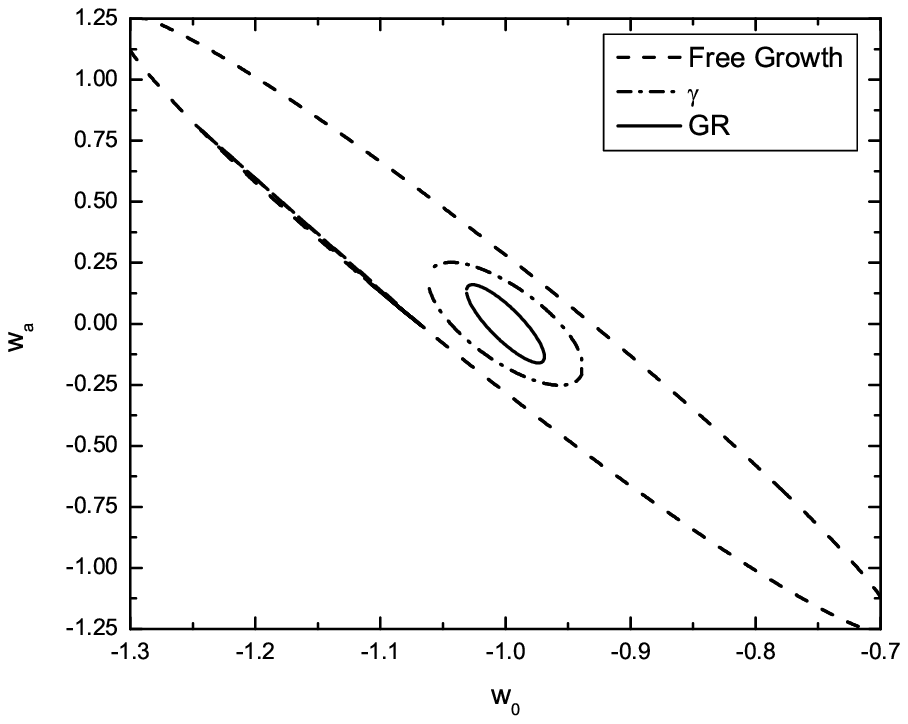}
   \includegraphics[width=0.75\textwidth]{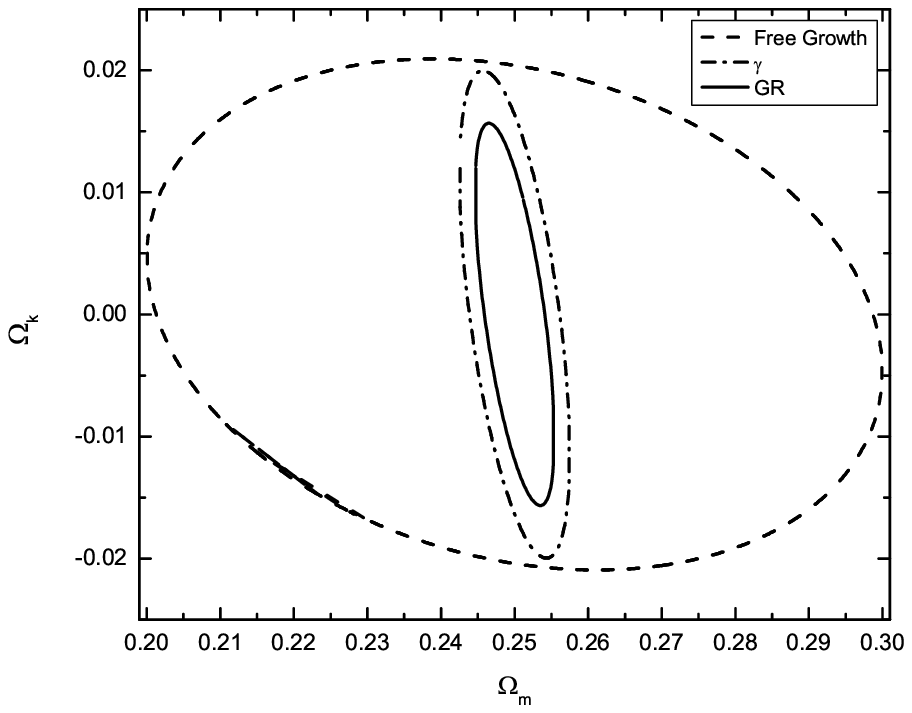}
    \caption{Upper panel shows one standard deviation confidence level 
     contours constraints on parameters $\omega_{a}$ and $\omega_{0}$ of 
     the $\omega$CDM parametrization, while lower panel shows these 
     for parameters $\Omega_{k}$ and $\Omega_{m}$.}
 \label{fig: 3}
\end{figure}

Figure \ref{fig: 3} shows constraints on parameters of the $\omega$CDM 
parametrization 
\citep[these should be compared to Figs.\ 4a and 5a of][]{samushia2010} .
When no assumptions are made about the nature of gravity the constraints 
on $\omega_0$ and $\omega_a$ are very weak and degenerate. When we assume 
general relativity the constraints tighten significantly, resulting in 
$\sim 10$\% accuracy in the measurement of $\omega_0$ and $\sim 25$\% 
accuracy in the measurement of $\omega_a$. 

\begin{figure}[H]
   \centering
   \includegraphics[width=0.75\textwidth]{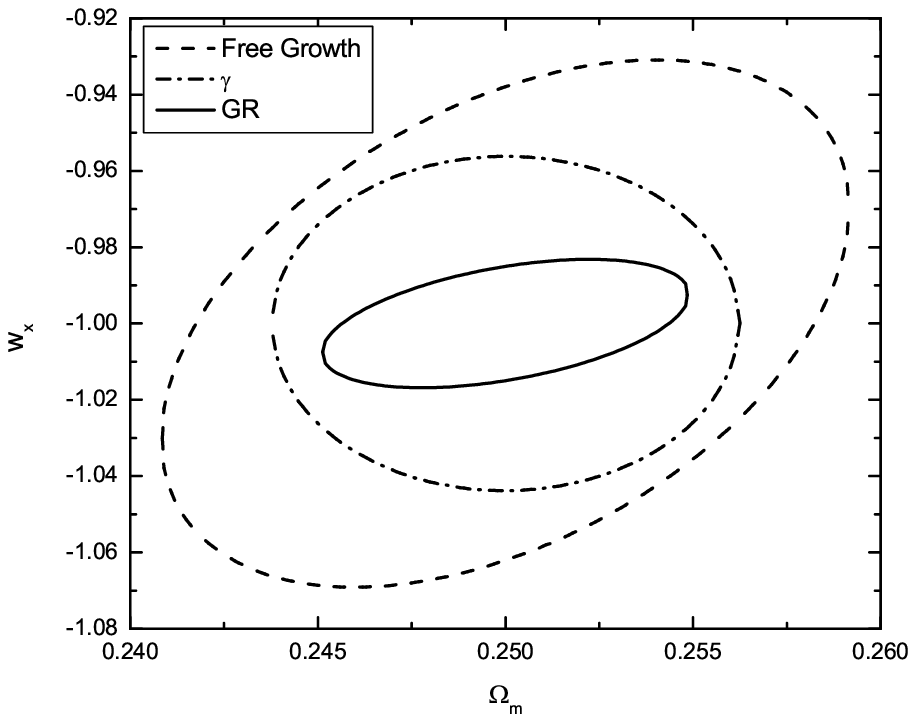}
   \includegraphics[width=0.75\textwidth]{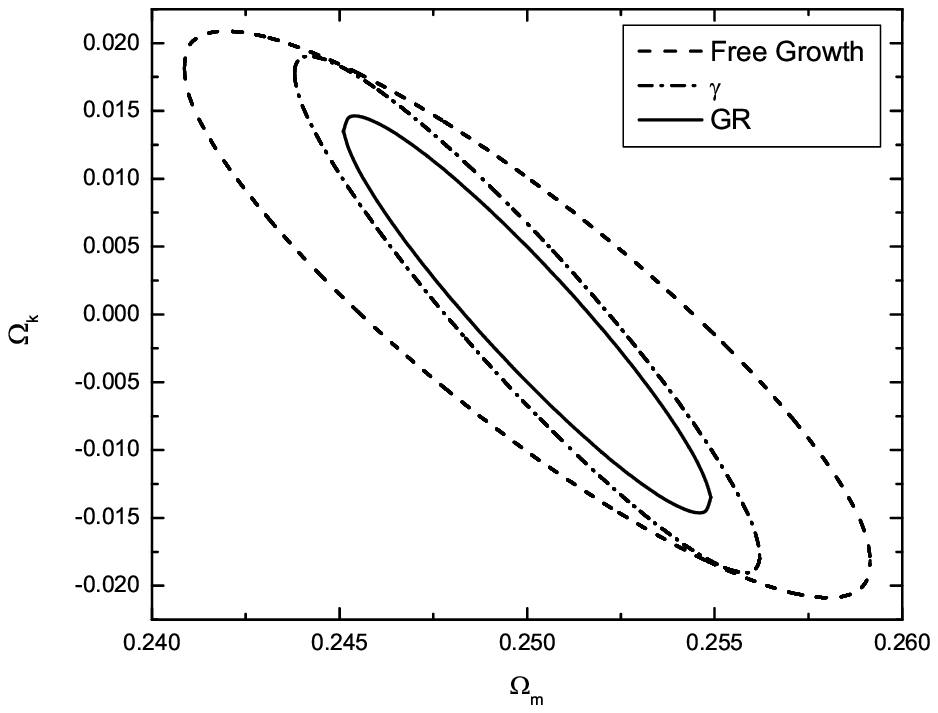}
   \caption{One standard deviation confidence level contour constraints on 
    parameters of the XCDM parametrization.}
 \label{fig: 4}
\end{figure}

The upper panel of Fig.\ \ref{fig: 4} shows constraints on the parameters 
$\omega_X$ and $\Omega_m$ of the XCDM parametrization. Similar to the 
previous case, the constraints tighten significantly when we assume 
general relativity as the model of gravity. About a 2\% measurement of 
$\omega_X$ and a 5\% measurement of $\Omega_m$ are possible in this case. 
The lower panel of Fig.\ \ref{fig: 4} show the related constraints 
on $\Omega_k$ and $\Omega_m$ for the XCDM parametrization. The 
constraints are similar to, but somewhat tighter than, those for the 
$\omega$CDM parametrization. This is because the XCDM parametrization 
has one less parameter than the $\omega$CDM parametrization. Spatial 
curvature can be constrained to about 15\% precision in this case.

\begin{figure}[H]
   \centering
   \includegraphics[width=0.75\textwidth]{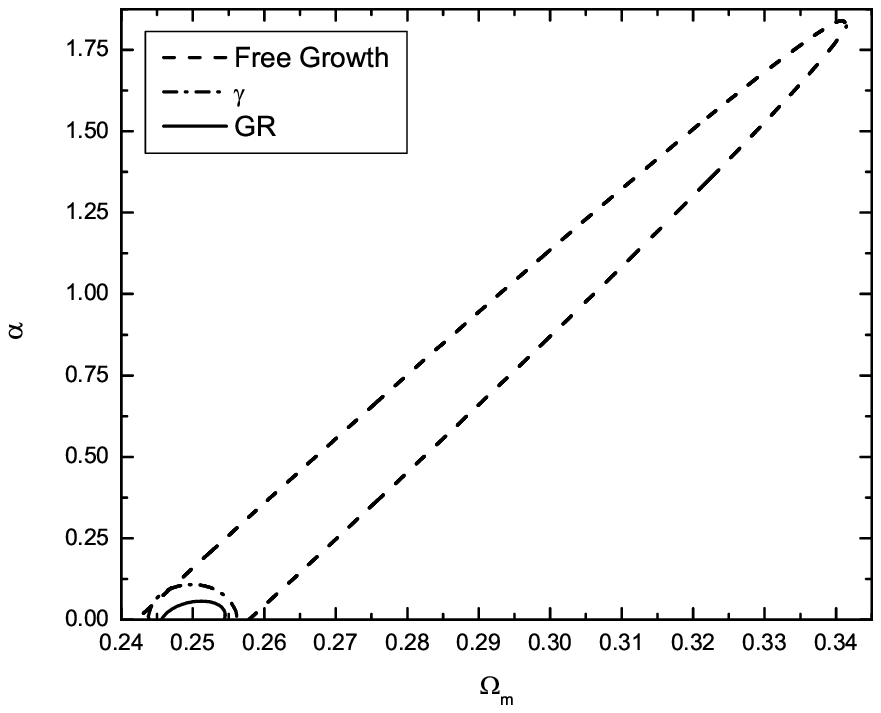}
   \includegraphics[width=0.75\textwidth]{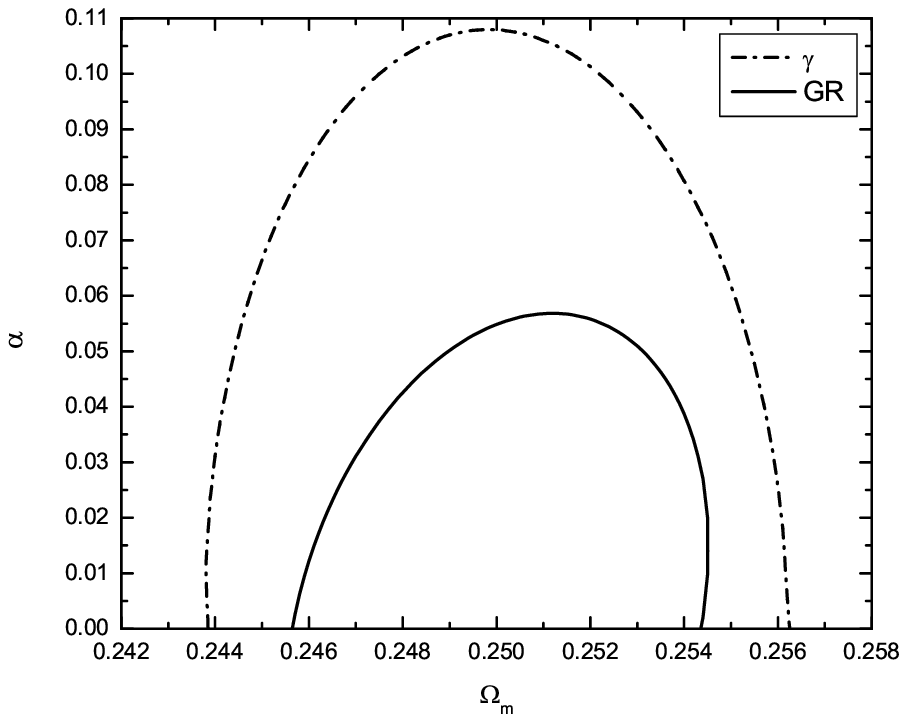}
    \caption{One standard deviation confidence level contour constraints 
     on parameters $\alpha$ and $\Omega_{m}$ of the $\phi$CDM model. Lower 
     panel shows a magnification of the tightest two contours in the 
     lower left corner of the upper panel.}
 \label{fig: 5}
\end{figure}

\begin{figure}[H]
   \centering
   \includegraphics[width=0.75\textwidth]{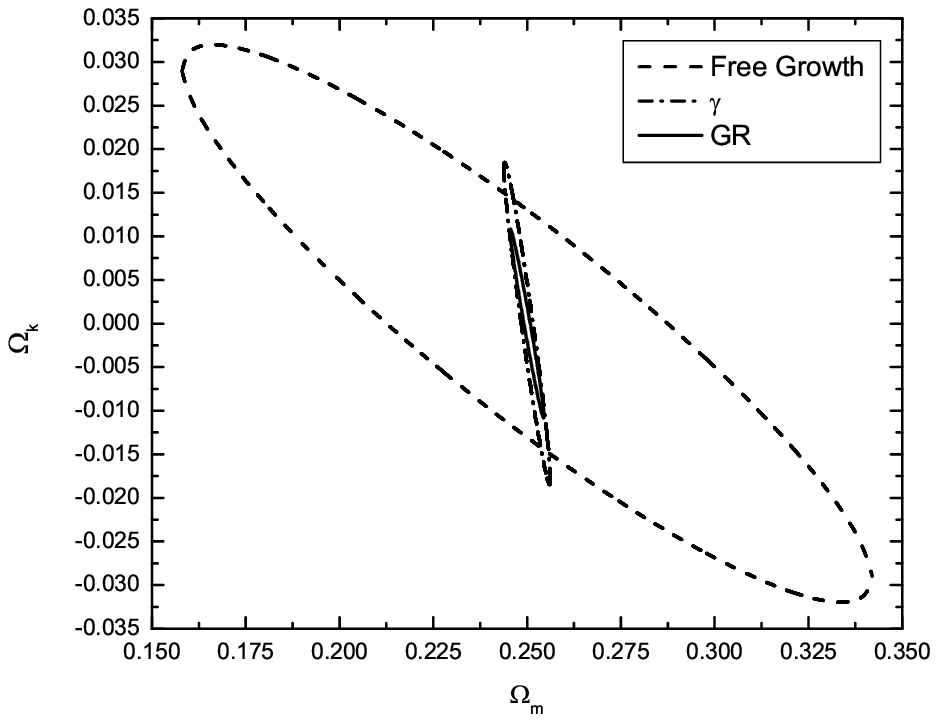}
   \includegraphics[width=0.75\textwidth]{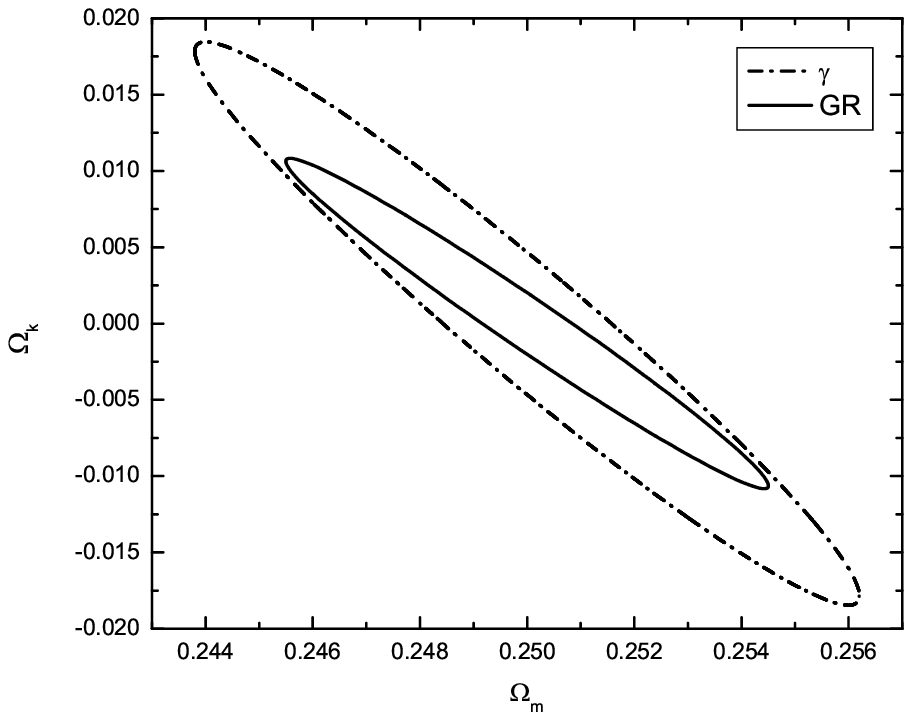}
   \caption{One standard deviation confidence level contour constraints on 
    parameters $\Omega_{k}$ and $\Omega_{m}$ of the $\phi$CDM model. The 
    lower panel shows a magnification of the two tightest contours in the 
    center of the upper panel.}
 \label{fig: 6}
\end{figure}

Figures \ref{fig: 5} and \ref{fig: 6} show constraints on parameters 
of the $\phi$CDM model. In the most general case, when no assumption 
is made about the nature of gravity, the constraints are weak and the 
parameters $\alpha$ and $\Omega_m$ are strongly correlated, with larger 
values of $\alpha$ requiring larger values of $\Omega_m$. When general 
relativity is assumed, the constraints become much stronger and
parameter $\alpha$ can be constrained to be less than 0.1 at the
1-$\sigma$ confidence level. This is significantly better than any
constraint available at the moment. 

\begin{figure}[H]
   \centering
   \includegraphics[width=0.8\textwidth]{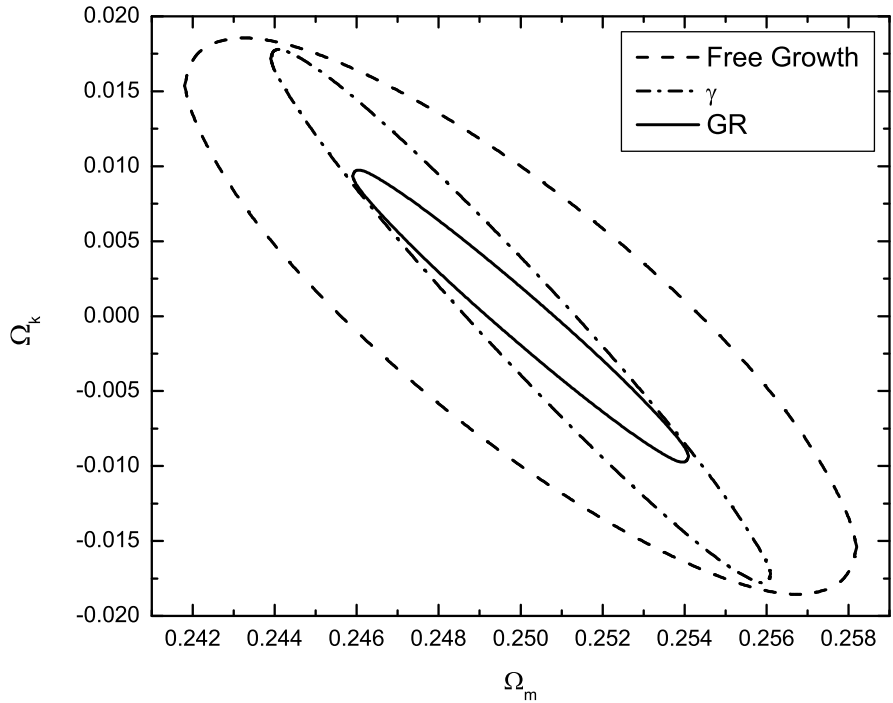}
    \caption{One standard deviation confidence level contour constraints on 
     parameters $\Omega_{k}$ and $\Omega_{m}$ of the $\Lambda$CDM model.}
 \label{fig: 7}
\end{figure}

Figure \ref{fig: 7} shows constraints on the parameters of the 
$\Lambda$CDM model. From the clustering data alone the spatial
curvature can be constrained with almost 1\% precision, largely 
because this model has the least number of free parameters.

The exact numerical values for the forecast error bars and likelihood contours
should be taken with caution and not be interpreted as predictions for the
performance of any specific survey (such as {\it Euclid} or {\it WFIRST}). Our
main objective in this work was first to investigate how the modified gravity
constraints change with different models of dark energy and second to
demonstrate the improvement in $\phi$CDM model constraints achievable with
future galaxy surveys. Because of this we were able to simplify our method by
adopting a Fisher matrix formalism instead of a full MCMC approach and also 
use a simplified description of the survey baseline. For more realistic 
predictions of {\it Euclid} performance, see, e.g., \citet{Laureijs2011, 
samushia2010, Majerotto2012}.

\section{Conclusion}
\label{sec:conclusion}

We have forecast the precision at which planned near-future space-based
spectroscopic galaxy surveys should be able to constrain the time dependence of
dark energy density. For the first time, we have used a consistent physical
model of time-evolving dark energy, $\phi$CDM, in which a minimally-coupled
scalar field slowly rolls down its self-interaction potential energy density.
We have shown that if general relativity is assumed, the deviation of the
parameter $\alpha$ of the $\phi$CDM model can be constrained to better than
$0.05$; this is almost an order of magnitude better than the best currently
available result.

The constraints on basic cosmological parameters, such as the relative 
energy densities of non-relativistic matter and spatial curvature, depend 
on the adopted dark energy model. We have shown that in the $\phi$CDM model 
the expected constraints are more restrictive than those derived using the
XCDM or $\omega$CDM parameterizations. This is due to the fact that the 
$\phi$CDM model has fewer parameters. Also, the XCDM and $\omega$CDM 
parameterizations assign equal weight to all possible values of $\omega$, 
while in the $\phi$CDM model there is an implicit theoretical prior on which 
equation of state parameter values are more likely, based on how easy it is
to produce such a value within the model.

Since the observational consequences of dark energy and modified gravity 
are partially degenerate, constraints on modified gravity parameters will 
depend on the assumptions made about dark energy. In Table \ref{tab: deviations} we 
show how the constraint on the $\gamma$ parameter depends on the adopted 
dark energy model. The constraints on $\gamma$ are most restrictive in the 
$\Lambda$CDM model. For the $\phi$CDM model the constraints on $\gamma$ are 
about a third tighter than those for the $\omega$CDM and XCDM parameterizations.

These results are very encouraging: data from an experiment of the type we
have modeled will be able to provide very good, and probably revolutionary,
constraints on the time evolution of dark energy. 

\acknowledgments
\section{Acknowledgments}
This work was supported by DOE grant DEFG030-99EP41093 and 
NSF grant AST-1109275. LS is grateful for support from European Research
Council, SNSF SCOPES grant \# 128040, and GNSF grant ST08/4-442.

\appendix
\section{Appendix}

In this Appendix we summarize how to estimate the precision of measurements
from the survey parameters.

The Fisher matrix coefficients are given by
\begin{equation}
F_{ij} = \frac{1}{2}\int_{k_{\rm min}}^{k_{\rm max}}\left(\frac{\partial\ln P}{\partial p^{i}} \right)\left(\frac{\partial\ln P}{\partial p^{j}} \right)V_{\rm eff}(k,\mu)\frac{d^{3} k}{(2\pi)^{3}} ,
 \label{equ:F_ij_apx}
\end{equation}
where the effective volume is
\begin{equation}
V_{\rm eff} = V_{0}\frac{nP(k,\mu)}{1 + nP(k,\mu)} , 
 \label{equ:V_eff}
\end{equation}
and $V_{0}$ is the total survey volume and $n$ is the number density.  Also,
following \citet{Tegmark:1997ab}, we multiply the integrand in Eq.\
(\ref{equ:F_ij_apx}) by a Gaussian factor $\exp\left(
-k^{2}\sigma_{z}\frac{dr(z)}{dz}\right)$, where $r(z)$ is the co-moving
distance, in order to account for the errors in distance induced by the errors
of redshift measurements, $\sigma_z = 0.001$. We model the theoretical power
spectrum using an analytic approximation of \citet{eisenstein98}. We integrate
in $k$ from $k_{\rm min} = 0$ to $k_{\rm max}$, where the $k_{\rm max}$ values
depend on redshift and are chosen in such a way that the small scales that are
dominated by non-linear effects are excluded. The range of scales that will be
fitted to the future surveys will depend on how well the theoretical templates
are able to describe small-scale clustering and is difficult to predict. The
$k_{\rm max}$ values along with the expected bias and number density of 
galaxies are listed in Table \ref{tab: FisherData}.

In order to derive the Fisher matrix of a specific cosmological model 
we have to go from our initial parameter space to the parameter space 
of the cosmological model whose Fisher matrix we want. The 
transformation formula for the Fisher matrix is given by 
\cite[see, e.g.,][for a review] {albrecht09}
\begin{equation}
F'_{lm} = \frac{\partial p_{i}}{\partial p'_{l}}\frac{\partial p_{j}}{\partial p'_{m}}F_{ij} ,
\label{equ:F_trans}
\end{equation}
where the primes denote the ``new" Fisher matrix and parameters. 

We now list the derivatives of the transformation coefficients of the 
$\phi$CDM model in the limit $\alpha \longrightarrow$ 0 and 
$\Omega_{k} \longrightarrow$ 0 (which corresponds to the fiducial 
spatially-flat $\Lambda$CDM model). The transformation coefficients 
relating $f_{\parallel}(z)$ and the parameters ($h, \Omega_{m}, 
\Omega_{k}, \alpha$) are
\begin{equation}
\frac{\partial f_{\parallel}(z)}{\partial h} = -\frac{1}{h} ,
\label{equ:fpar_h}
\end{equation}
\begin{equation}
\frac{\partial f_{\parallel}(z)}{\partial \Omega_{m}} = \frac{1}{2E(z)^{2}}[1 - (1 + z)^{3}] ,
\label{equ:fpar_Om}
\end{equation}
\begin{equation}
\frac{\partial f_{\parallel}(z)}{\partial \Omega_{k}} = \frac{1}{2E(z)^{2}}[1 - (1 + z)^{2}] ,
\label{equ:fpar_Ok}
\end{equation}
\begin{equation}
\frac{\partial f_{\parallel}(z)}{\partial \alpha} = -\frac{(1 - \Omega_{m})}{8E(z)^{2}} .
\label{equ:fpar_alpha}
\end{equation}
For the other transformation coefficients, it is convenient to 
introduce the integral
\begin{equation}
D(z) = \int_{0}^{z}\frac{dz'}{E(z')} .
\label{equ:int_Ez}
\end{equation}
Then the transformation coefficients between $f_{\perp}(z)$ and the
parameters ($h, \Omega_{m}, \Omega_{k}$, $\alpha$) are 
\begin{equation}
\frac{\partial f_{\perp}(z)}{\partial h} = -\frac{1}{h} ,
\label{equ:fper_h}
\end{equation}
\begin{equation}
\frac{\partial f_{\perp}(z)}{\partial \Omega_{m}} = \frac{1}{2D(z)}\int_{0}^{z}\frac{dz'}{E(z')^{3}}[1 - (1 + z')^{3}],
\label{equ:fper_Om}
\end{equation}
\begin{equation}
\frac{\partial f_{\perp}(z)}{\partial \Omega_{k}} = \frac{D(z)^{2}}{6} + \frac{1}{2D(z)}\int_{0}^{z}\frac{dz'}{E(z')^{3}}[1 - (1 + z')^{2}],
\label{equ:fper_Ok}
\end{equation}
\begin{equation}
\frac{\partial f_{\perp}(z)}{\partial \alpha} = -\frac{(1 - \Omega_{m})}{8D(z)}\int_{0}^{z}\frac{dz'}{E(z')^{3}}.
\label{equ:fper_alpha}
\end{equation}

Finally, the transformation coefficients between the growth factor $f(z)$ and 
the parameters ($\gamma, h, \Omega_{m}, \Omega_{k}, \alpha$) are
\begin{equation}
\frac{\partial f(z)}{\partial \gamma} = \frac{f(z)}{\gamma}\ln f(z) ,
\label{equ:f_gamma}
\end{equation}
\begin{equation}
\frac{\partial f(z)}{\partial \Omega_{m}} = \frac{\gamma f(z)}{\Omega_{m}E(z)^{2}}\left\lbrace E(z)^{2} - \Omega_{m}[(1 + z)^{3} - 1]\right\rbrace ,
\label{equ:f_Om}
\end{equation}
\begin{equation}
\frac{\partial f(z)}{\partial \Omega_{k}} = -\frac{\gamma f(z)}{E(z)^{2}}[(1 + z)^{2} - 1] ,
\label{equ:f_Ok}
\end{equation}
\begin{equation}
\frac{\partial f(z)}{\partial \alpha} = -\frac{\gamma f(z)}{4E(z)^{2}}[1 - \Omega_{m}] .
\label{equ:f_alpha}
\end{equation}

\begin{table}[H]
\caption{Values of the $k_{max}$, bias $b(z)$ from \citet{Orsi:2010ab}, and the number densities $n(z)$ taken from \citet{Geach:2010ab}.}
\centering
\begin{tabular}{|c|c|c|c|}
  \hline
  $z$ & $k_{max}$ & $b(z)$ & $n(z)$ \\
  \hline
  0.55 & 0.144 & 1.0423 & 3220 \\
  \hline
  0.65 & 0.153 & 1.0668 & 3821 \\
  \hline
  0.75 & 0.163 & 1.1084 & 4364 \\
  \hline
  0.85 & 0.174 & 1.1145 & 4835 \\
  \hline
  0.95 & 0.185 & 1.1107 & 5255 \\
  \hline
  1.05 & 0.197 & 1.1652 & 5631 \\
  \hline
  1.15 & 0.2 & 1.2262 & 5972 \\
  \hline
  1.25 & 0.2 & 1.2769 & 6290 \\
  \hline
  1.35 & 0.2 & 1.2960 & 6054 \\
  \hline
  1.45 & 0.2 & 1.3159 & 4985 \\
  \hline
  1.55 & 0.2 & 1.4416 & 4119 \\
  \hline
  1.65 & 0.2 & 1.4915 & 3343 \\
  \hline
  1.75 & 0.2 & 1.4873 & 2666 \\
  \hline
  1.85 & 0.2 & 1.5332 & 2090 \\
  \hline
  1.95 & 0.2 & 1.5705 & 1613 \\
  \hline
  2.05 & 0.2 & 1.6277 & 1224 \\
  \hline
\end{tabular}
 \label{tab: FisherData}
\end{table}


\begin{table}[H]
\caption{Predicted deviations of parameter $\gamma$ from its fiducial value,
 at one standard deviation confidence level, for different assumptions about 
 dark energy.}
\centering
\begin{tabular}{|c|c|c|}
  \hline
  DE model & Fiducial $\gamma$ & deviation \\
  \hline
  $\omega$CDM & 0.55 & 0.035 \\
  $\phi$CDM & 0.55 & 0.023 \\
  XCDM & 0.55 & 0.035 \\
  $\Lambda$CDM & 0.55 & 0.016 \\
  \hline
\end{tabular}
 \label{tab: deviations}
\end{table}
\end{document}